\documentstyle[aps,floats,epsf,psfig,twocolumn]{revtex}
\begin{document}
\tighten
\draft 
\title{Quasiparticle Inelastic Lifetime from Paramagnons in
Disordered Superconductors}
\author{T. P. Devereaux}
\address{Department of Physics, University of Waterloo, 
Waterloo, ON N2L 3G1, Canada}
\address{~
\parbox{14cm}{\rm 
\medskip
	The paramagnon contribution to the quasiparticle
inelastic scattering rate in disordered superconductors is presented.
Using Anderson's exact eigenstate formalism, it is shown that the
scattering rate is Stoner enhanced and is further enhanced by the
disorder relative to the clean case
in a manner similar to the disorder enhancement of the long-range
Coulomb contribution.  The results are discussed in connection with the
possibility of conventional or
unconventional superconductivity in the borocarbides. The results are
compared to recent tunneling experiments on
LuNi$_{2}$B$_{2}$C.
}}
\maketitle
\narrowtext
\section{Introduction}

In the early seventies, certain rare earth ternary compounds
were found to display superconductivity
while at the same time showed strong tendencies to be magnetic. A large
body of theoretical work has been devoted to the interplay of magnetism
and superconductivity\cite{review}. Recently, there is
increasing evidence that there
is an interplay of magnetism and superconductivity
in the boro-carbides\cite{boro}
as well as RuSr$_{2}$GdCu$_{2}$O$_{8}$\cite{ru}.
Presently it is widely debated whether these materials are
conventional superconductors with sharply peaked density of states (DOS) 
near the Fermi level (similar to some A-15's), or unconventional
with ground state pairing of lower symmetry than the underlying lattice.
For instance, in $R$=Lu, Y 
$R$Ni$_{2}$B$_{2}$C, scanning tunneling microscopy (STM)
has given evidence for conventional BCS behavior, albeit with a substantially
smeared DOS, with a smearing parameter $\Gamma/\Delta=0.2$\cite{wilde}. 
Optical conductivity studies also support moderately
strong coupled conventional superconductivity with $2\Delta/T_{c}=3.9-5.2$
\cite{optics}. At the same time de Haas-van Alphen\cite{dehaas},
magnetic field anisotropy\cite{maki},
and electronic Raman scattering\cite{raman} experiments
have given evidence for at least very small 
gaps over a portion of the Fermi surface.

Substitutional or positional disorder has played a crucial role in determining
whether a material is conventional (i.e., obeys Anderson's theorem) or not,
and recently studies on borocarbides doped with Co have been 
performed\cite{cheon}.
Heat capacity and magnetic measurements on 
Y(Ni$_{2-x}$Co$_{x}$)B$_{2}$C\cite{hoell}
have interpreted the drop in T$_{c}$ with increasing Co doping as due to the 
reduction
of the DOS at the Fermi level rather than pairbreaking by nonmagnetic 
impurities.
On the other hand, Raman measurements on the same systems
have shown an increase in spectral weight below the
gap edge as Co is doped in, contrary to conventional BCS behavior\cite{raman}.

However it is well known that conventional superconductors which are highly
disordered display substantially smeared BCS properties which can mimic
unconventional pairing\cite{dynes}. This can result
from vanishing of phase coherence or from the interplay of interactions and
disorder. The latter is most manifest in the reduction of quasiparticle (qp)
lifetimes. Inelastic collisions broaden qp 
eigenstates and lead to a
smearing of activated or threshold behavior in single- and two-particle 
correlation functions, measured e.g., by
tunneling, optical conductivity, and electronic Raman scattering. 
While the present status of the superconducting ground state
of the borocarbides remains unclear, it is of interest to inspect
whether strong inelastic scattering can modify s-wave properties to
the point where the ubiquitous exponential behavior of various
thermodynamic and transport quantities is obscured.  For instance, 
the absence of a coherence peak in NMR is usually interpreted as a signal of
unconventional electronic pairing. However, it is well known
that the coherence peak can be suppressed as a consequence of strong
inelastic electron-phonon collisions \cite{fibich}. 
While the coherence peak can be
fully suppressed only for large electron-phonon couplings in clean
superconductors \cite{allen}, it has been shown that the
peak can be
further suppressed in disordered superconductors and is absent in
the region of strong disorder for only moderate couplings \cite{nmr}.

The microscopic interplay of disorder, magnetic fluctuations
and superconductivity is reflected in
the behavior of the qp inelastic lifetime.
In this paper we present a calculation for the qp
inelastic scattering rate due to spin fluctuations within a formalism
developed in previous works \cite{all}.  The calculation is undertaken
by first obtaining an effective fluctuation propagator in the
superconducting state, and then using the exact eigenstate formalism
as used in the case of Coulomb scattering with the replacement of the
Coulomb propagator and vertex with the derived fluctuation propagators
and vertices. It is shown that the rate is qualitatively similar to the
rate due to Coulomb interactions with addition of the Stoner
enhancement. Finally we discuss our results in terms of STM data on
the borocarbide superconductors.

\section{Calculations}

The
scattering rate from paramagnons in clean superconductors on a lattice
is well known for the case of $s-$ or $d-$wave
superconductors\cite{quinlan}. 
The calculation for the inelastic scattering rate due to paramagnon
exchange in disorder metals is also well known\cite{abrahams}. 
In both cases the results are similar to the 
scattering rate from
long-range Coulomb interactions, with an additional enhancement via the Stoner
factor $1/(1-I)$, where $I=UN_{F}$, $U$ is a phenomenological short range 
interaction, and $N_{F}$ is the DOS per spin at the Fermi level.
For dirty metals and superconductors, the electron-phonon interaction is
reduced via ``collision drag'' relative to the clean case\cite{schmid}, while
the electron-electron interaction
is enhanced by disorder due to the breakdown of screening by diffusive
electrons\cite{schmid2,all}. 
The latter enhancement of the scattering rate at the Fermi
surface is $\hat \rho^{3/2} (E_{F}/T)^{1/2}$ compared
to that of 3D clean materials. Here $\hat \rho$ is the dimensionless measure of
disorder, with $\hat\rho=\rho/\rho_{M}$, where
$\rho$ is the extrapolated residual resistivity and $\rho_{M}$ the
Mott number, which in a jellium model is given by
$\rho_{M}=3\pi^{2}/e^{2}k_{F}$. We use units such that $k_{B}=\hbar=1$. 
However, calculations for the scattering rate calculated for
superconductors on a
lattice\cite{quinlan} have treated impurities and interactions
independently and therefore do not capture the disorder enhancement
derived for conventional superconductors. Therefore in this paper we
investigate the interplay of
disorder, superconductivity, and magnetism by revisiting the problem of
inelastic scattering.

The spin fluctuation propagator is given by the sum of 
longitudinal $K_{\uparrow \uparrow}$ and transverse 
$K_{\uparrow\downarrow}$ paramagnons, respectively\cite{berk}. 
They can be expressed in
terms of the polarization bubble $\chi$ as 
\begin{eqnarray}
K_{\uparrow \uparrow}=&U\chi U + U \chi U \chi K_{\uparrow
\uparrow}, \nonumber \\
K_{\uparrow \downarrow}=&U + U \chi K_{\uparrow \downarrow}.
\end{eqnarray}
Solving these equations we obtain the fluctuation propagator 
$t({\bf q},\omega)$ 
\begin{equation}
t({\bf q},\omega)=K_{\uparrow \uparrow}+K_{\uparrow \downarrow}=
{U^{2}\chi({\bf q},\Omega)\over{1-U^{2}\chi^{2}({\bf q},\Omega)}}+
{U\over{1-U\chi({\bf q},\Omega)}}.
\end{equation}
However in the superconducting state one must distinguish between
charge and spin response couplings due to their different coherence
factors. Therefore in the superconducting state the
propagator splits into two contributions given by\cite{paobickers}
\begin{eqnarray}
t_{c}({\bf q}, \Omega)= {1\over{2}}
{U^{2}\chi_{c}({\bf q},\Omega)\over{1+U\chi_{c}({\bf q},\Omega)}},\\
t_{s}({\bf q}, \Omega)= {3\over{2}}
{U^{2}\chi_{s}({\bf q},\Omega)\over{1-U\chi_{s}({\bf q},\Omega)}}-
U^{2}\chi_{s}({\bf q},\Omega),
\nonumber
\end{eqnarray}
with $\chi_{c,s}$ the charge, spin susceptibilities, respectively.
In the following we perform calculations in the continuum limit and neglect
lattice effects. This is certainly important in order to capture strong
scattering via qp exchange of antiferromagnetic reciprocal lattice vector
momenta ${\bf Q}$. However, the incipient magnetic instability via paramagnon
exchange nevertheless is reflected via the Stoner criterion. Albeit a naive
approach to the borocarbides or other materials with strong antiferromagnetic
fluctuations, the results allow us to qualitatively estimate the effects of
disorder on qp inelastic scattering from paramagnons.

The gauge-invariant charge polarization $\chi_{c}$ has been calculated in 
disordered superconductors in Ref. \cite{gang}.  It has the structure
$\chi({\bf q},\omega) = B({\bf q},\omega) + B_{C}({\bf q},\omega)$.
Here $B$ is the density response function in the pair approximation
\cite{prange}, while $B_{C}$ contains the collective excitation (the
Anderson-Bogolubov mode) which restores gauge invariance. 
It was shown that for $k_{F}\xi \gg 1$ 
collective effects can be ignored and that the "pair
approximation" for the polarization is adequate, where 
$\xi=\sqrt{1/m\pi\Delta}\hat\rho^{-1}$ is
the dirty-limit coherence length, 
For $T=0$ the polarization can be written as
\begin{eqnarray}
\chi_{c}^{\prime\prime}({\bf q},\omega) = 
\phi^{\prime\prime}({\bf q},\sqrt{\omega(\omega-2\Delta)}) 
\Theta(\omega-2\Delta) \nonumber \\
\times
\left[(\omega+2\Delta)E(\alpha)-{4\Delta\omega\over{\omega+2\Delta}}K(\alpha)\right],
\end{eqnarray}
while the spin susceptibility is given by
the Mattis-Bardeen result\cite{mb}
\begin{eqnarray}
\chi_{s}^{\prime\prime}({\bf q},\omega) = \phi^{\prime\prime}({\bf
q},\sqrt{\omega(\omega-2\Delta)}) \Theta(\omega-2\Delta) \nonumber \\
\times
\left[(\omega+2\Delta)E(\alpha)-4\Delta K(\alpha)\right].
\end{eqnarray}
Here, $\alpha={\omega-2\Delta\over{\omega+2\Delta}}$, and $E$ and $K$
are complete elliptical integrals of the first and second kinds,
respectively. $\phi^{\prime\prime}$ is the spectrum of the density
Kubo function for noninteracting electrons. It can calculated by a
variety of techniques for various limits of disorder.  For clean
metals, the spectrum is white,
\begin{equation}
\phi^{\prime\prime}({\bf q}, \epsilon)={m^{2}\over{4\pi q}},
~~~~~~~~~~\rm{clean},
\end{equation}
while for diffusive qp dynamics, $\phi^{\prime\prime}$ is
given by a diffusion pole
\begin{equation}
\phi^{\prime\prime}({\bf q},\epsilon)=N_{F}
{Dq^{2}\over{(Dq^{2})^{2}+\epsilon^{2}}},~~~~~~~~~~\rm{diffusive},  
\end{equation}
with $D$ the diffusion constant. Here we have neglected Cooper propagator
renormalization, which can be shown to give a smaller contribution to
the scattering rate than Diffusion
propagator renormalization by a factor of $1/k_{F}\xi$.

The limiting behavior for finite temperatures with
$T<< \Delta $ is given as:
\begin{eqnarray}
\chi^{\prime\prime}_{c,s}({\bf q}, \Omega << \Delta) &\approx&
\Omega \phi^{\prime\prime}({\bf q},\sqrt{2\Delta\Omega}) e^{-\Delta/T}
\nonumber \\
&&\times \cases{1,~~~~~~ \rm{charge}, \cr
(\Delta/T)\ln(4T/\Omega),~~~~~ \rm{spin},} \\
\chi^{\prime\prime}_{c,s}({\bf q}, \Omega \ge 2\Delta) &\approx&
\Delta \phi^{\prime\prime}({\bf q},2\sqrt{2} \Delta) \sqrt{\pi T/\Delta}
e^{-\Delta/T} \nonumber \\
&&\times \cases{ 1/2,~~~~~ \rm{charge}, \cr
1,~~~~~ \rm{spin}.}
\end{eqnarray}
Thus the behavior of the spin and charge susceptibilities
yields different contributions to the paramagnon scattering
in the charge and spin channel.

The paramagnon contribution to the self energy can be split
in the usual way into an anomalous and even and odd normal pieces.
It has been shown for the case of long-range
Coulomb interactions that the even
part of the normal self energy contribution can be ignored, and can be
shown for the spin-fluctuation case as well\cite{all}. Expanding near the qp
pole in the BCS Green's function\cite{kaplan}, we obtain the
expression for the on-shell inelastic scattering rate due to paramagnons
exchange in the charge channel $\Gamma_{c}$ and spin channel
$\Gamma_{s}$,
\begin{eqnarray}
&&\Gamma_{c,s}(\omega)= - {1\over{Z^{\prime}}} \sum_{{\bf q}} \int
{d\epsilon\over{\pi N_{F}}} \phi^{\prime\prime}({\bf q},
\epsilon-\omega) \\
&&\times \int_{0}^{\infty} {dx\over{\pi}} [f(x+\omega)+n(x)]
t^{\prime\prime}_{c}({\bf q},x) \nonumber \\
&&\times\left[G^{\prime\prime}(\epsilon,\omega+x)\pm\Delta/\omega
F^{\prime\prime}(\epsilon,\omega+x)\right] + (\omega \rightarrow -\omega),
\nonumber
\end{eqnarray}
where $n, f$ are Bose and Fermi distributions, respectively,
$G^{\prime\prime}$ and $F^{\prime\prime}$ are the imaginary parts of
the bare normal and anomalous BCS Green's functions, respectively,
$Z^{\prime}$ is the real part of the qp renormalization, and
$(\omega \rightarrow -\omega)$ denotes the addition of terms which
differ from the ones written only by the sign of $\omega$.

Substituting Eqs.(3-9) into Eq.(10), we obtain the inelastic
scattering rate $\tau^{-1}_{s}=2\Gamma_{s}$. It can be shown that the 
contribution to the scattering rate from the charge
channel yields a subdominant contribution for all values of disorder and
interaction $I<1$ compared to the long-range Coulomb contribution
calculated in Ref. \cite{all}. Therefore for the remainder of the paper
we neglect $\Gamma_{c}$ and focus on $\Gamma_{s}$.
The scattering rate
is dominated by qp population at the gap edge.
For $T=0$, an injected qp must have enough energy to give up to 
break a Cooper pair ($3\Delta$) and for   
${\Omega-3\Delta\over{\Omega+3\Delta}} << 1$ we obtain,
\begin{eqnarray}
\Gamma_{s}^{T=0}(\Omega\ge 3\Delta)&&= 
{3I^{2}\pi^{2}\over{16(1-I)^{3/2}}} {\Delta\over{Z^{\prime}}}
{\Delta\over{\Omega}}{\Delta\over{E_{F}}}F\left({\Omega\over{\Delta}}\right)
\nonumber\\
&&\times\cases{ {1\over{\sqrt{1-I}}},~~~~~~~ \rm{clean},\cr
\left({3\hat\rho\over{\pi}}\right)^{3/2}
\sqrt{{E_{F}\over{\Delta}}}, ~~~~~\rm{dirty,}}
\end{eqnarray}
with $F(x)=x(x^{2}/2-x-1)\sqrt{(x-2)^{2}-1} 
+(x/2-2)\ln[x-2+\sqrt{(x-2)^{2}-1}]$.
At finite temperatures, the
Cooper pair recombination
rate is dominated by the kinematic factor $\Gamma^{R}
\propto e^{-2\Delta/T}$
and qp scattering rate $\Gamma^{S}\propto e^{-\Delta/T}$.
For a qp at the gap edge, the dominant contribution to the
recombination rate is given by
\begin{eqnarray}
\Gamma_{s}^{R}(\Delta >> T) = && {3\pi^{2}I^{2}\over{8\sqrt{2}(1-I)^{3/2}}} 
{T^{2}\over{Z^{\prime}E_{F}}}
e^{-2\Delta/T}\nonumber \\
&&\times \cases{ {1\over{\sqrt{1-I}}}, ~~~~~\rm{clean,}\cr
\left({3\hat\rho\over{\pi}}\right)^{3/2}
\sqrt{{E_{F}\over{\Delta}}}, ~~~~~\rm{dirty,}}
\end{eqnarray}
while for the scattering rate we obtain to leading order
\begin{eqnarray}
\Gamma_{s}^{S}(\Delta >> T) = && {3\pi I^{2}\ln(2)\over{8(1-I)^{3/2}}} 
{\Delta\over{Z^{\prime}}}\sqrt{{\pi T\over{2\Delta}}}
e^{-\Delta/T}\nonumber \\
&&\times \cases{ {1\over{\sqrt{1-I}}}, ~~~~~\rm{clean,}\cr
\left({3\hat\rho\over{\pi}}\right)^{3/2}
\sqrt{{E_{F}\over{\Delta}}}, ~~~~~\rm{dirty.}}
\end{eqnarray}
We see a similar behavior between the paramagnon and long-range
Coulomb contributions to the inelastic scattering
rate\cite{all}. $\Gamma_{s}$ possesses the same temperature
dependence as the Coulomb contribution, with the exponential
temperature dependence reflecting the necessity of two qps
per scattering event. Further, we see the same disorder enhancement 
($\hat \rho^{3/2} (E_{F}/\Delta)^{1/2}$ ) relative to the clean
case as in the long-range Coulomb case. Lastly, we note that 
the energy
gap $\Delta$ acts as a cut off for the divergence of the rate that
occurs in the $2-d$ dirty normal calculation \cite{abrahams}, just as
in the long-range Coulomb case\cite{all}.

On top of the disorder enhancement, there is the Stoner
enhancement relative to the Coulomb contributions
due to the nearness of a magnetic instability. In
materials close to the instability, this contribution will be dominant
over the Coulomb and phonon terms except for very low temperatures,
where the power-law temperature dependence of the phonon contribution
takes over \cite{all,kaplan}. 
We note that our expression are valid for $\Delta/E_{F}<< 1-I<< 1$,
i.e., provided
that one is not too close to the Stoner criterion for magnetism, $I=1$.
At the instability, the rate
saturates as it does in the case of a normal metal near the
metal-insulator transition \cite{wyso}.  
However in order to accurately describe the
dynamics at the magnetic transition one needs to use a more
sophisticated spin fluctuation propagator than the one derived here
from RPA diagrams only, which tend to overestimate paramagnon effects
\cite{naka}. 

Finally, we can compare the results to the values of the scattering rates
inferred from STM data on clean and thin films of LuNi$_{2}$B$_{2}$C. 
To our knowledge, a temperature dependence
of the scattering rate has not yet been published, nor has a reliable
estimate of the scattering rate been made from optical (Raman or infrared)
or Hall probes as has been done in the high T$_{c}$ cuprates. Moreover
no systematic study of the effects of impurities and doping have been made
concerning the scattering rate. Nevertheless we can estimate if 
inelastic scattering from paramagnons in an s-wave superconductor
is sufficient to explain the broadening observed in STM
measurements\cite{wilde}. As a rough estimate for $1/\tau_{s}$ we take
$I \sim 2/3$, $Z^{\prime}=1/2$,
Fermi velocity $v_{F}\sim 3.5\times 10^{7}$
cm/s, Fermi energy $E_{F}\sim 0.3$ eV given from LDA estimates for
LuNi$_{2}$B$_{2}$C from Ref. \cite{Pickett}.
STM data taken at low temperatures
in Ref. \cite{wilde} gives $\Delta=18$cm$^{-1}$, which is consistent with
Raman measurements\cite{raman}. This yields a scattering rate for clean
systems at $T=0.5T_{c}$ from Eq. (13) of $1/\tau_{s}=
1.3\times 10^{-3}$meV, or $1/\tau_{s}\Delta=6 \times 10^{-4}$, which is
clearly too small to match experiments.  
Either the scattering is most likely due to electron-phonon
collisions\cite{estimate} or perhaps due to large gap anisotropy. 

Since $\rho(T=0)$ increases
quickly as Co is doped in\cite{cheon}, rising by over an order of
magnitude for 15\% Co doping\cite{wilde}, it may be feasible that
the disorder enhancement for $1/\tau_{s}$ in an s-wave
scenario could lead to spectral
weight at low frequencies observed via magnetic field anisotropy\cite{maki}
or Raman\cite{raman} measurements. An estimate for the Mott number 
is difficult since the parameters $v_{F}, k_{F}$ and the other parameters
entering in Eqs. (12-13) are presumably
disorder dependent, 
and it is not clear where the metal-insulator transition
occurs for this compound. A conservative estimate from the Ioffe-Regel
criterion in Ref. \cite{wilde} gives $\rho_{M}\sim 400 \mu\Omega$-cm.
Therefore taking $\rho(T=0) \sim 100 \mu\Omega-$cm as in Ref. \cite{wilde}
into Eq. (13) only gives $1/\tau_{s}\Delta\sim10^{-3}$, which is clearly
too small to account for the large broadening observed via STM even in
relatively clean films nor is it sufficient to account for the substantial
spectral weight observed at low frequencies via Raman scattering. It is
tempting to therefore conclude that the large broadening comes either from
nodal qps in
conventional (extended s-) or unconventional (d-)
pair states. 

However, there are problems in each scenario.
Small amount of Co doping (on the few percent level) quickly push these
materials into the dirty limit ($\xi/l << 1$)\cite{cheon}. 
If the gap possessed extended $s-$wave symmetry, 
the disorder would be sufficient to wash
out any remaining anisotropy and necessarily lead to sharp threshold
behavior, which is not observed. 
On the other hand, the disorder would also lead to a sharp drop
in $T_{c}$ if the gap possessed $d-$wave symmetry and 
unconventional 
superconductivity would be expected to be completely suppressed\cite{big}
for 15\% Co doping, which again is not observed.
Therefore it is unclear from current data whether superconductivity
is conventional or not, and perhaps the situation is clouded by the
presence of additional
non-superconducting bands, which would also yield a non-vanishing
zero bias conductance and low frequency spectral weight.
It would thus be extremely
useful to study impurity and cation
dopings further to determine if the enhanced scattering rates are
responsible for the behavior indicative of unconventional pairing as the
disorder is increased. Raman scattering measurements would be very useful in
this regard, and remains a topic for further investigation.

\end{document}